\documentclass[journal]{IEEEtran}
\ifCLASSINFOpdf
\else
\fi

\usepackage{url}
\usepackage{mathtools}
\usepackage[round]{natbib}

\usepackage{cuted}
\title{Beyond Correlation: A Path-Invariant Measure for Seismogram Similarity}

\author{Joshua Dickey, Brett Borghetti, William Junek, and Richard Martin
\thanks{J. Dickey, B. Borghetti and R. Martin are with the Department of Electrical and Computer Engineering, Air Force Institute of Technology, Wright-Patterson AFB, OH 45433 USA e-mail: joshuadickey@gmail.com.}
\thanks{W. Junek is with the Air Force Technical Applications Center, Patrick AFB}
}
\date{April 2019}

\begin{document}

\maketitle

\begin{abstract}
Similarity search is a popular technique for seismic signal processing, with template matching, matched filters and subspace detectors being utilized for a wide variety of tasks, including both signal detection and source discrimination. Traditionally, these techniques rely on the cross-correlation function as the basis for measuring similarity. Unfortunately, seismogram correlation is dominated by path effects, essentially requiring a distinct waveform template along each path of interest. To address this limitation, we propose a novel measure of seismogram similarity that is explicitly invariant to path. Using Earthscope's USArray experiment, a path-rich dataset of 207,291 regional seismograms across 8,452 unique events is constructed, and then employed via the batch-hard triplet loss function, to train a deep convolutional neural network which maps raw seismograms to a low dimensional embedding space, where nearness on the space corresponds to nearness of source function, regardless of path or recording instrumentation. This path-agnostic embedding space forms a new representation for seismograms, characterized by robust, source-specific features, which we show to be useful for performing both pairwise event association as well as template-based source discrimination with a single template.
\end{abstract}

\section{INTRODUCTION}

\begin{figure}[htbp]
\centering
\includegraphics[width=3in]{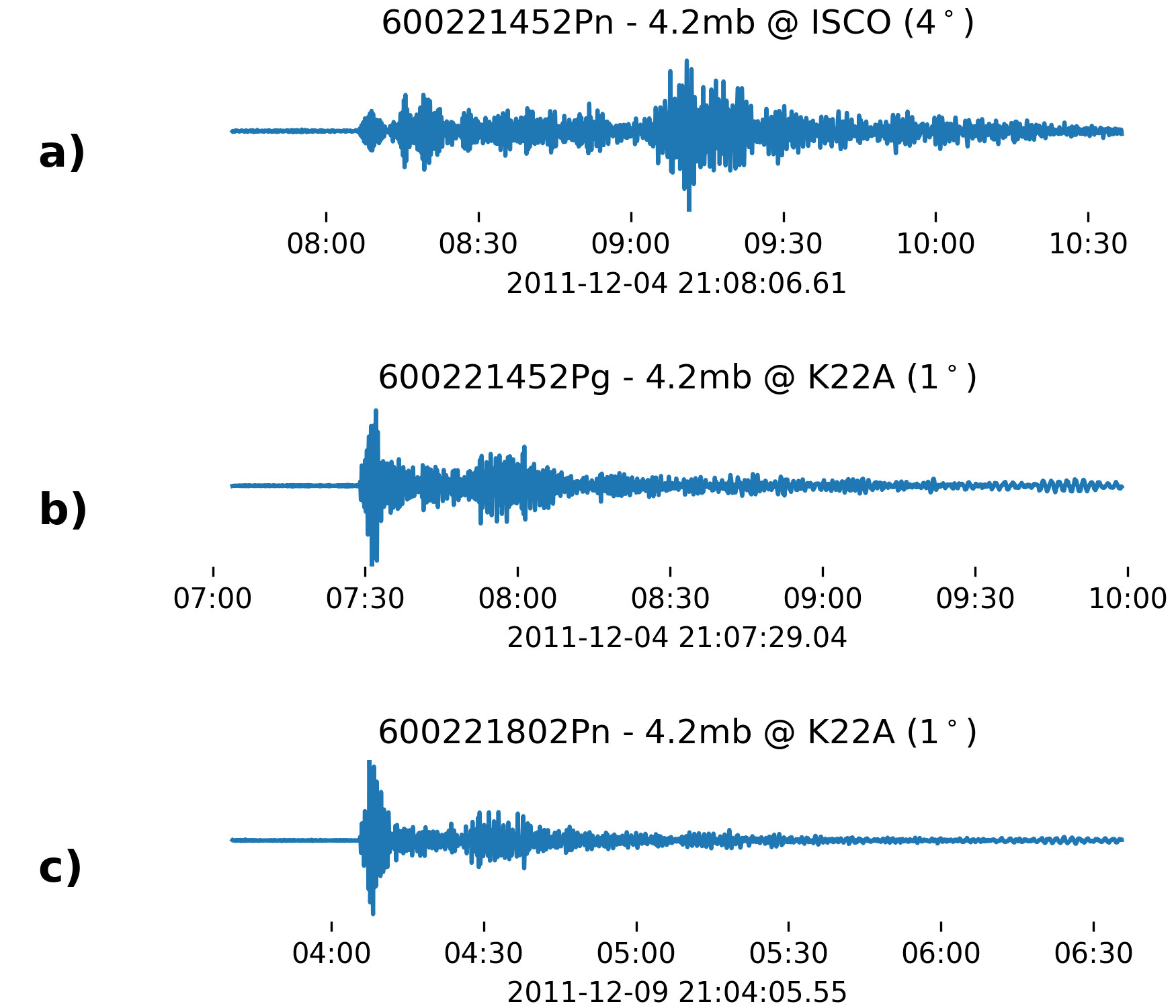}
\caption[Seismograms of Three Explosions near Thunder Basin]{Three Seismograms depicting explosions at a coal mine near Thunder Basin, WY. Seismograms a) and b) depict a common source event (600221452), recorded at two separate seismic stations, ISCO and K22A respectively. Seismogram c) depicts a nearby event (600221802), also recorded at K22A. Seismograms a) and b) depict the same event recorded at different stations, while seismograms b) and c) depict different events which share a common path. The correlation between the same-source waveforms a) and b) is only 0.03, and the waveforms visually appear quite different. On the other hand, the visual similarity between the path-similar waveforms b) and c) is obvious, and they are correlated with a coefficient of 0.18.  This illustrates the path-dominant similarity inherent to seismogram correlation.}
\label{fig:SeisSim}
\end{figure}

\IEEEPARstart{S}{eismograms} are time-series records of the earth's motion at a fixed station. This motion results from seismic waves, that have often traveled a considerable distance from the source event, and seismograms reflect the combined influence of both the source itself, and the propagation path between source location and recording station~\citep{Bormann2002a}.  As illustrated in Fig.~\ref{fig:SeisSim}, two seismograms depicting different events yet sharing a common path can appear similar. This fact has long been recognized by the seismic community~\citep{Stauder1967, Kanamori1978}. In the earliest days of manual processing and helicorders, analysts were often able to identify mining events from a particular mine, recorded at a particular station, by simply comparing the visual similarity of new seismograms to previously recorded examples~\citep{Israelsson1990}. In fact, a common practice was to take two translucent paper seismograms and compare them, by passing the waveforms across one another while holding them up to a light source~\citep{Schulte1993}. Thus began the science of seismogram similarity. Of course, the advent of computer processing ushered in the development of a multitude of techniques to exploit these similarities algorithmically. Case-based discrimination~\citep{Dysart1987}, template matching~\citep{Giannakis1990}, waveform correlation~\citep{Harris1991}, subspace detection~\citep{Harris2006} and similarity search~\citep{Yoon2015} are all similarity-based algorithms which have been proposed over the last several decades, and deployed against a wide range of seismic signal processing tasks, such as discriminating mining blasts, screening swarm events, identifying aftershock sequences, and even detecting general seismic signals.

While these algorithms have different tasks ranging from discrimination to detection, fundamentally they are all examples of similarity-based classifiers~\citep{Chen2009}, which estimate the class label of a new seismogram based on its similarity to one or more previously labeled templates. Furthermore, these similarity-based classifiers all share a common measure of similarity: cross-correlation. Such methods are generally referred to as correlation detectors~\citep{Harris2006}. 

This common reliance on correlation is concerning, because the correlation coefficient of two seismograms is dominated by path effects~\citep{Schulte1993}, as demonstrated in Fig.~\ref{fig:SeisSim}. While path-dominant similarity can be desirable, such as when detecting aftershock sequences from a particular fault, or mining blasts from within a small quarry, in general, path-dominant similarity is problematic, as source-similar signals de-correlate with even slight deviations in path~\citep{Harris2006}. This includes deviations in origin location, such as two explosions occurring at different points in a mining quarry, and deviations in recording location, such as two recordings of the same explosion by separate seismic stations in a regional seismic array. In either case, path differences of even just a quarter wavelength can significantly degrade the correlation of two seismograms \citep{Pechmann1982, Motoya1985}.

\begin{figure}[t]
  \centering
  \setlength\fboxrule{0pt}
  \fbox{\includegraphics[width=3in]{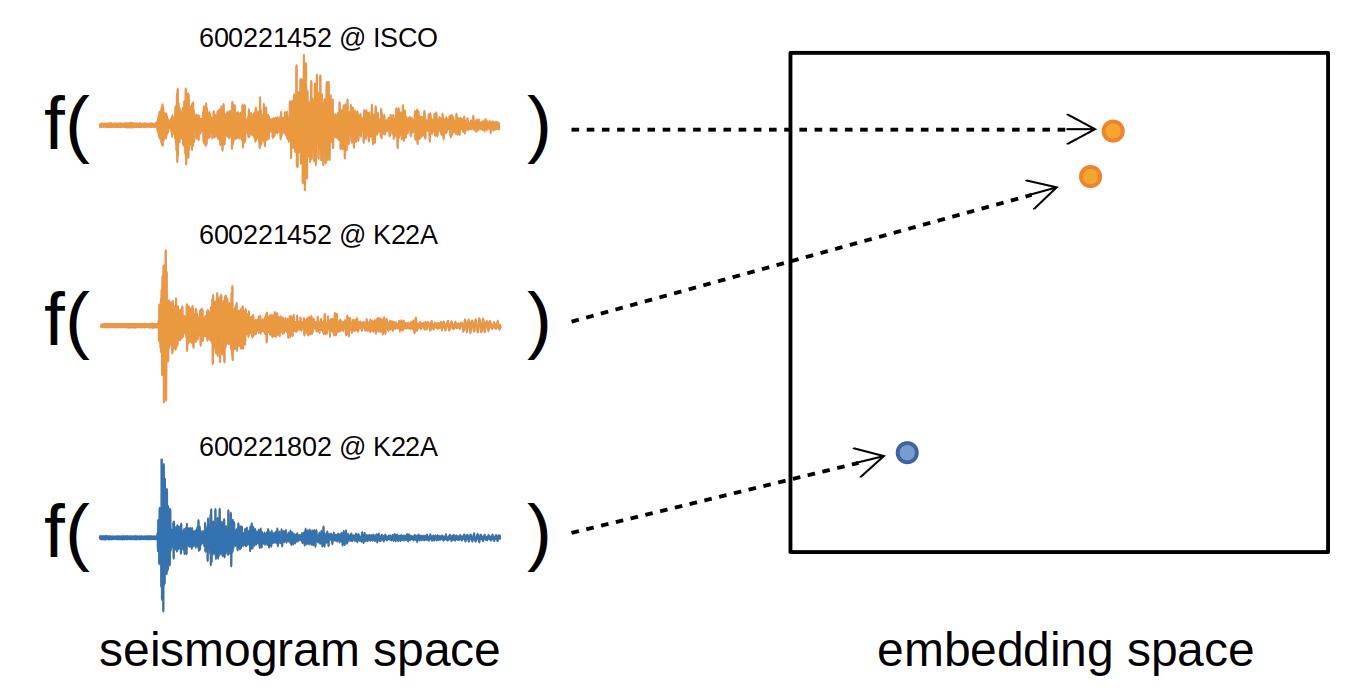}}
\caption[Path-Invariant Embedding Function for Seismograms]{Path-Invariant Embedding Function for Seismograms. The embedding function, $f(\cdot)$, is a non-linear transformation that maps time-series seismograms to low-dimensional embeddings. The mappings should be path-invariant and source-specific, such that regardless of the recording station, all seismograms associated with a particular event are mapped closely in the embedding space, and seismograms not associated with that event have more distant embeddings, as demonstrated in this notional diagram. This embedding function can be learned using a convolutional neural network architecture, trained with seismogram triplets.}
\label{fig:Embed}
\end{figure}

This work presents a new measure for seismogram similarity that bypasses correlation entirely, and that is designed to be both path-invariant and source-specific. 
To be precise, the design goal is to create a measure of seismogram similarity that enables the identification of seismograms sharing a common source event, regardless of the path of travel. While such a measure was previously computationally intractable, it is possible with the careful application of deep convolutional neural networks (CNNs). In 2019, researchers at the Los Alamos National Laboratory published a method using a CNN to predict the pairwise association of seismic phase arrivals, for 6 second windows, across a local group of 6 stations in northern Chile, reporting an accuracy of over 80\%~\citep{McBrearty2019}. Building on these results, we construct a source-dominant, path-invariant measure for seismogram similarity which operates on 180 second windows and is generalized across more than 1,000 sensors across North America. We do this by utilizing a state-of-the-art machine learning technique from the field of facial recognition, called a Triplet Network, which not only indicates pairwise association between seismograms, but actually maps the seismograms to low-dimensional vectors, called embeddings, such that the embedding space distance between seismograms sharing a common source event are minimized, regardless of path, while remaining distinct from any other events. This embedding strategy is displayed in Fig.~\ref{fig:Embed}. In this way, the embedding function becomes a rich feature extraction technique for source-specific and path-invariant features.

\begin{figure*}[htbp]
  \centering
  \setlength\fboxrule{0pt}
  \fbox{\includegraphics[width=6in]{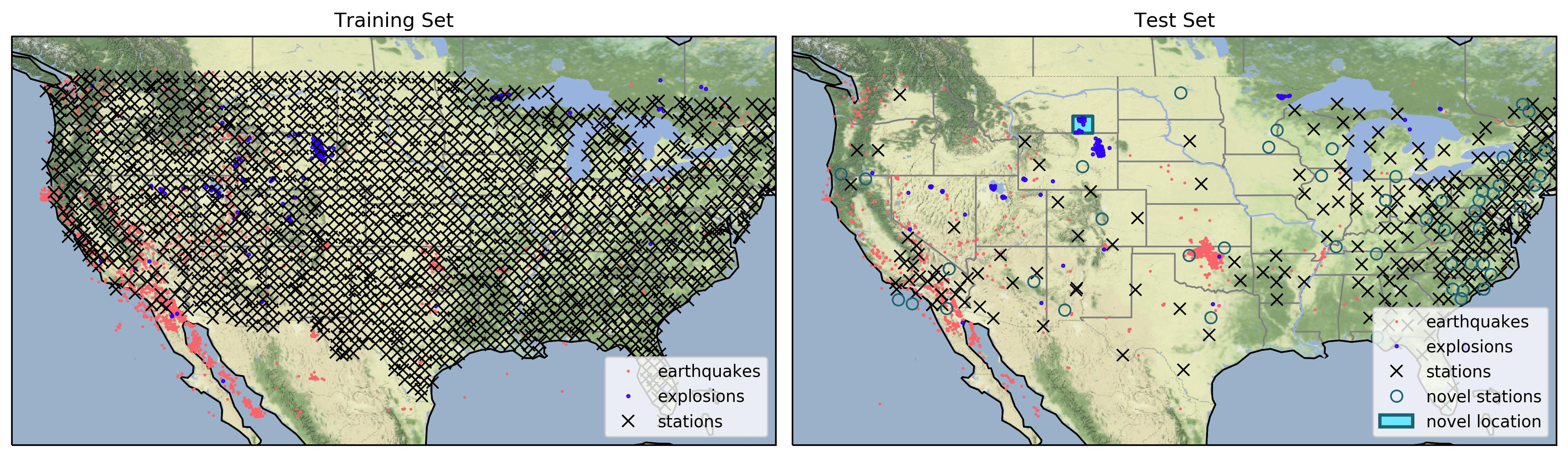}}
\caption[USArray Station Map]{Map showing the geographical location of each recording station and event in the training and testing datasets. The majority of the stations were installed as part of the Earthscope's Transportable USArray, and were in operation from 18 to 24 months before being moved. Additionally, 51 novel stations and a small region of novel event locations are unique to the test set.}
\label{fig:SourceMap}
\end{figure*}

The triplet network architecture accepts three observations - two similar and one different from the others. Training a triplet network to learn seismic source similarity requires source-similar seismogram triples: two of the three waveforms are associated with a common source event and the third waveform is not. For this task, it is preferable to have a training set containing seismograms recorded from a densely-spaced sensor network, so that the neural network can experience seismograms recordings across numerous paths for the same event. The 400 three-channel broadband sensors of the USArray experiment provided an ideal dataset of seismograms;  data from this array is used for training and testing.  The triplet network is trained against 13 years of data (2007 - 2013), validated against a single year of data (2014), and tested against the final two years of data (2015-2016). Additionally, a subset of 51 recording stations and a small region of event locations was held out from the algorithm during training, to allow a proper evaluation of the generalizability of the technique. A map detailing the dataset is shown in Fig.~\ref{fig:SourceMap}.

The value of this path-invariant measure is demonstrated through performance evaluation on two common seismic tasks: event association, and source discrimination. The event association task of determining whether or not two waveforms depict the same event achieves a binary accuracy of 80\%. This accuracy is achieved using only the waveform characteristics, without information on times or recording locations, and the technique has strong potential to augment existing methods of event association~\citep{McBrearty2019}. 

The real promise of the technique, however is for source discrimination. The embedding space is a rich basis for source-specific seismic feature extraction~\citep{Hadsell2006b}. Our similarity-based explosion discriminator achieves 95.8\% accuracy with no explicit training for the source discrimination task; the discriminator simply compares the similarity of unknown waveforms to a single randomly-selected explosion template. This technique is often referred to as one-shot learning~\citep{Koch2015}, and shows promise for discrimination of novel sources when only a few extant templates are available.

In the remainder of this work, these contributions and conclusions are explored in detail, by reviewing the related literature, outlining methodology, and detailing and discussing our results. 

\section{BACKGROUND}

This work merges two relatively disparate fields of science. On the one hand, the application is seismogram similarity, a field with a rich history and considerable previous research. On the other hand, the methodology employs learned similarity, a relatively nascent field that has principally been associated with machine learning image processing applications. This background section is divided into three distinct subsections: seismogram similarity; learned similarity; and learned seismogram similarity. Each subsection contains a brief background and literature review, as well as a discussion of the limitations and gaps in the current research, which our work attempts to fill.

\subsection{Seismogram Similarity}

A seismogram represents the composition of several effects, including the seismic source itself, the propagation path from the source to the seismometer, the frequency response of the seismometer, as well as any ambient noise at the seismometer's location~\citep{Bormann2002a}. Because of this diverse composition, estimating and even defining seismogram similarity can be quite challenging.

The traditional measure for seismogram similarity is the cross-correlation function. This measure has been used for detecting and discriminating seismic signals since the late 1980s~\citep{Dysart1987}, and such techniques are commonly referred to as correlation detectors~\citep{Harris2006}. Correlation detectors are exquisitely sensitive, allowing detections near the noise floor for known repeating events in highly confined geographical regions~\citep{Gibbons2006}. Unfortunately, this confinement is also a limitation, as seismogram correlation has been shown to decay exponentially with even minor differences in path distance~\citep{Israelsson1990}. In fact, early research suggested that correlation-based similarity was limited to signals with hypo-centres separated by no more than a quarter wavelength~\citep{Frankel1982, Motoya1985}, although later efforts have since shown improvements, allowing the correlation length to be up to two wavelengths~\citep{Harris1991}. Additionally, researchers have also shown that seismograms quickly decorrelate across small variations in mechanism and source function~\citep{Hutchings1990}. These facts limit the applicability of the correlation detector to only the most repetitive sources that are confined to localized geographical regions~\citep{Harris2006}. 

To increase the applicability of the correlation detector, there have been numerous adaptations proposed. To address variations in ambient noise, narrow bandpass filters were applied~\citep{Israelsson1990}. To address minor variations in mechanism, composite templates were employed, derived from linear combinations of several master templates representing a range of mechanisms~\citep{Harris1991}. To address path effects, dynamic waveform matching was developed, introducing a non-linearity to the correlation, allowing relative stretching or squeezing of the template~\citep{Schulte1993}. Subspace detectors attempt to address all of these variations at once, with even more robust composite templates~\citep{Harris2006}. Recently, efforts focused on a multiplicity of templates and a computationally efficient search across them~\citep{Yoon2015, Zhang2015, Beauce2017, Bergen2018}. These efforts have significantly increased the effectiveness of correlation-based detectors. In fact, for regions with a high sensor density, such as Northern California, it is estimated that more than 90\% of events have sufficient similarity to be detected via correlation~\citep{Waldhauser2008}. However, this figure is highly dependent on both the density of the sensor network and the completeness of the template library~\citep{Tibi2017}. As such,
Dodge and Walter estimate that still only 18\% of all global events possess sufficient similarity to be detected by these methods~\citep{Dodge2015}.

In summary, cross-correlation is a powerful measure for seismogram similarity, especially as a tool for detecting highly-repeating path-specific events. However, cross-correlation is fundamentally limited as a general measure of seismogram similarity, due to its inherent path-dependence. In this study, we address this limitation directly, and propose an alternative measure of seismogram similarity that is invariant to path, instrumentation and ambient noise.

\subsection{Learned Similarity}

Each of the traditional seismogram similarity measures discussed so far has been fundamentally built around the cross-correlation function. However, it is interesting to note that almost none of those measures performed cross-correlation directly on the raw waveforms. Instead, each measure first applied some pre-processing function to the raw waveforms, either linear (time shifts, bandpass filters, linear combinations) or non-linear (dynamic time warping) prior to performing cross-correlation. We can generally understand these pre-processing functions to be mappings, from raw waveform space to a new \textit{embedding space}. In each case, the mapping function is chosen such that the cross-correlation of two objects in the embedding space meets some desired similarity objective. 

As it turns out, this embedding process used in traditional correlation-based similarity closely mirrors the process accomplished in machine learning-based similarity. For learned similarity, a parameterized embedding function architecture is established, and the parameters are optimized such that the distance between two objects in the space achieves the desired similarity objective. Over the last several years, such learned similarity measures have revolutionized the field of facial recognition in particular and the field of image processing in general, fueling advances in image recognition~\citep{Wang2014}, object tracking~\citep{Leal-Taixe2016b} and even vision navigation~\citep{Kumar2016}. In the remainder of this section, we review some of the state of the art techniques available for constructing deep learned similarity measures, focusing particularly on the embedding function architecture and similarity objective, in turn. 

\subsubsection{\textit{Embedding Function Architecture}}

Many early efforts to create learned similarity spaces utilized a linear architecture, such as the Mahalanobis distance~\citep{Xing2002, Jain2008, Jain2009}. However, in recent years, much success has been gained by employing non-linear architectures~\citep{Belkin2003}, particularly in the form of deep convolutional neural networks or CNNs~\citep{Hadsell2006b}. These CNNs were originally developed with 2-dimensional kernels, or filters, which allowed them to closely model the hand-crafted kernels traditionally used in image processing~\citep{Lecun1989}. To adapt these powerful CNN architectures to process time-series waveforms, 1-dimensional CNNs were developed~\citep{Burges2003}, enabling learned similarity spaces for audio waveforms~\citep{Jang2009}.

A more recent advancement to the traditional CNN architecture is the Temporal Convolutional Network (TCN), which is characterized by layered stacks of dilated causal convolutions and residual connections~\citep{Bai2018}, as illustrated in Fig.~\ref{fig:TCN_stack_SeisSim}. Such an architecture is  particularly applicable to time-series waveforms with long-period dependencies, and offers several distinct advantages for seismic feature extraction~\citep{Dickey2019}, including: 
\begin{itemize}
\item Residual connections allow the model to have high-capacity and stable training.
\item Dilated convolutions allow precise control over the receptive field.
\end{itemize}

The receptive field is of primary importance for time-series modeling, as it explicitly limits the learnable feature periodicity at a given layer. The equation for calculating the receptive field, $r$, for a given convolutional layer, $l$, kernel size $k$, and dilation rate, $d$ is given in \eqref{eq:receptiveField_sim}:

\begin{equation}
\begin{aligned}
&r_l = r_{l-1} + d(k - 1) \\
&\text{where $r_0 = 0$}
\label{eq:receptiveField_sim}
\end{aligned}
\end{equation}

In summary, the TCN is ideally suited for the efficient embedding of seismograms. This architecture presents a rich search space for learning an optimal embedding function. However, optimizing this function requires defining a suitable similarity objective, detailed next.

\subsubsection{\textit{Similarity Objective}}

Defining a quantitative similarity objective begins with a qualitative understanding of what similarity means for the given task, which is often referred to as a semantic definition of similarity. Once the semantic definition is established, the next step is to approximate it with an embedding function, such that nearness in the embedding space implies the semantic similarity~\citep{Chopra2005b}. This embedding function is learned via back-propagation of loss, $\mathcal{J}$, that reinforces the semantic definition.


One of the simplest semantic definitions of similarity is the concept of a \textit{match}, where a matched pair of objects share a common identity, and an unmatched pair objects have different identities. For example, in the facial recognition task, a matched pair is defined as two images of the same person and an unmatched pair is defined as two images of distinct persons. The similarity objective is to optimize the parameters of the embedding function such that the embedding space distance between matched pairs is small, while the distance between unmatched pairs is large. This embedding function can be learned directly by a Siamese Neural Network, which takes in a batch of $m$ object pairs, of which half are matched, and half are unmatched. The two objects, $X_A^{(i)}$ and $X_B^{(i)}$, are then embedded via twin copies of the embedding function, $f(\cdot)$, with tied parameter weights $w$. The parameters of the embedding function are updated via the contrastive loss function, which penalizes two contrasting cases: matched pairs are penalized for being embedded too far apart and non-matched pairs are penalized for being embedded too close together with respect to some margin, $\alpha$, as given in Eq. \eqref{eq:match_loss} and Eq. \eqref{eq:nonmatch_loss}, respectively~\citep{Chopra2005b}. 

\begin{equation}
\begin{aligned}
 \mathcal{J} = \sum^{m/2}_{i=1} \left[ \Big \langle  f(X_A^{(i)}) , f(X_B^{(i)}) \Big \rangle \right]
\label{eq:match_loss}
\end{aligned}
\end{equation}

\begin{equation}
\begin{aligned}
 \mathcal{J} = \sum^{m/2}_{i=1} \left[ \alpha - \Big \langle  f(X_A^{(i)}) , f(X_B^{(i)}) \Big \rangle \right] \small_+ \\
\text{where [ ]$_+$ indicates the ramp function.} 
\label{eq:nonmatch_loss}
\end{aligned}
\end{equation}

This technique works well, however, one drawback is the relatively inefficient use of the embedding space. Matches are too greedy, as the Siamese Network attempts to map all matches to a single point in the space. Meanwhile, non-matches are inefficient, pushed apart only a fixed distance~\citep{Hoffer2015b}. As a result, the Siamese Network is used less frequently in favor of the Triplet Network, which we shall next discuss.

 The Triplet Network is similar to the Siamese Network~\citep{Hoffer2015b}, however it is trained on batches of $m$ triples, where each triple is comprised of an anchor object, $X_A^{(i)}$, a positive object, $X_P^{(i)}$, and a negative object, $X_N^{(i)}$. From within each triple, both a matched and non-matched pair can be constructed, however, the triplet loss function computes the relative embedding distance between the matched pair and non-matched pair, and no loss is accrued as long as the matched pair is closer by some margin, $\alpha$, as given in Eq. \eqref{eq:TripletLoss}. 

\begin{equation}
\begin{aligned}
 \mathcal{J} = \sum^{m}_{i=1} \left[ \Big \langle  f(X_A^{(i)}) , f(X_P^{(i)}) \Big \rangle - \Big \langle f(X_A^{(i)}) , f(X_N^{(i)}) \Big \rangle + \alpha \right] \small_+ \\
 \text{where [ ]$_+$ indicates the ramp function.} 
 \label{eq:TripletLoss}
 \end{aligned}
\end{equation}

The Triplet network avoids the greediness of the Siamese network, and makes more efficient use of the embedding space, however it has its own drawbacks. Particularly, it can converge quickly at first, but learning slows rapidly, as the majority of the negative pairs are pushed beyond the margin, failing to train the weights appreciably. This can be solved by sampling hard pairs, semi-hard pairs and several other sampling strategies, all of which rely on iterative processing via forward propagation to determine embedding space distances, selectively sampling based on those distances, and then applying back propagation on the sample~\citep{Hermans2017}. The algorithm used to sample hard pairs is commonly referred to as the \textit{batch hard} loss function, and it requires that each batch be composed by randomly sampling $L$ distinct identities  and then randomly sampling $K$ examples of each identity. In this way, the total number of objects in a batch is $L*K$, and each object is double indexed so that object $X_u^{(v)}$ represents the $u_{th}$ example of the $v_{th}$ identity. The triplet loss is calculated using Eq. \eqref{eq:TripletLoss}, except that in this case, every object in the batch is treated as an anchor $X_A^{(i)}$, and used to form a new triplet by selecting the \textit{hardest} positive and \textit{hardest} negative samples, $X_P^{(i)}$ and $X_N^{(j)}$ respectively, for that anchor within that batch, as detailed in Eq.~\eqref{eq:BatchHardLoss}.

\begin{strip}
\begin{equation}
\begin{multlined}
 \mathcal{J} = \overbrace{ \sum^{L}_{i=1} \sum^{K}_{A=1} }^{\text{all anchors}} \Bigg[ 
  \overbrace{
     \max_{\begin{subarray}{l} P=1...K \\ P \neq A \end{subarray}}
     \Big\langle  f(X_A^{(i)}) , f(X_P^{(i)}) \Big\rangle
 }^{\text{hardest positive}} - 
 \overbrace{
     \min_{\begin{subarray}{l} j=1...L \\ N=1...K \\ j \neq i \end{subarray}} \Big\langle f(X_A^{(i)}) , f(X_N^{(j)}) \Big\rangle 
 }^{\text{hardest negative}}
 + \alpha \Bigg] \small_+ \\
  \text{where [ ]$_+$ indicates the ramp function.} 
 \label{eq:BatchHardLoss}
\end{multlined}
\end{equation}
\end{strip}

\subsection{Deep Seismogram Similarity}
Deep Neural Networks are now being used across many areas of seismological research, from earthquake detection to earthquake early warning systems, ground-motion prediction, seismic tomography, and even earthquake geodesy~\citep{Kong2018}. However, no effort has been made to date to use deep neural networks to build a seismogram similarity metric. The closest related work was in early 2019, where researchers at Los Alamos National Labs published a paper describing a convolutional neural network for the pairwise association of seismograms depicting a common event, regardless of path~\citep{McBrearty2019}. This work shows that path-invariant features do exist within the seismogram record. The seismograms considered in their work had a signal length of 6 seconds, and were restricted to recordings from 6 seismic stations.  To process the signals, they used a shallow CNN with 4 layers, the input accepting two seismograms, the output producing a single boolean. This results in a similar output to a Siamese network, but without tied weights. The lack of tied weights means there is no embedding layer, which prevents their technique from being used for feature extraction. And the small number of stations limits the generalizability and transportability of their algorithm.  Finally, the short signal length (6 s) limits each individual seismogram to containing a single phase arrival, thereby limiting its ability to extract long-period features, such as P and S wave energy ratios, which are particularly pertinent to general source discrimination tasks. 

\section{METHODOLOGY}

  We present a novel seismogram similarity measure, based on a learned embedding function, that is both source-dominant and path-invariant. We show that the resultant embedding space is a rich representation space for seismic signals, useful for performing similarity-based classification against two common class dichotomies for seismograms: common event vs different events (event association)  and earthquake vs explosion (source discrimination). The remainder of this section describes the embedding function architecture, the similarity objective, the USArray dataset and the evaluation criteria for the two classification tasks.

\subsection{Embedding Function Architecture}

The goal is to learn a path-invariant embedding function for seismograms, useful for source discrimination at up to regional distances. This is accomplished using a hybrid architecture with two distinct parts: First, a TCN is employed with a receptive field wide enough to capture both P and S wave phases; second, a densely connected output layer, with 32 nodes, is employed to facilitate a rich low-dimensional embedding space. 

\begin{figure}[htbp]
  \centering
  \setlength\fboxrule{0pt}
  \fbox{\includegraphics[width=3in]{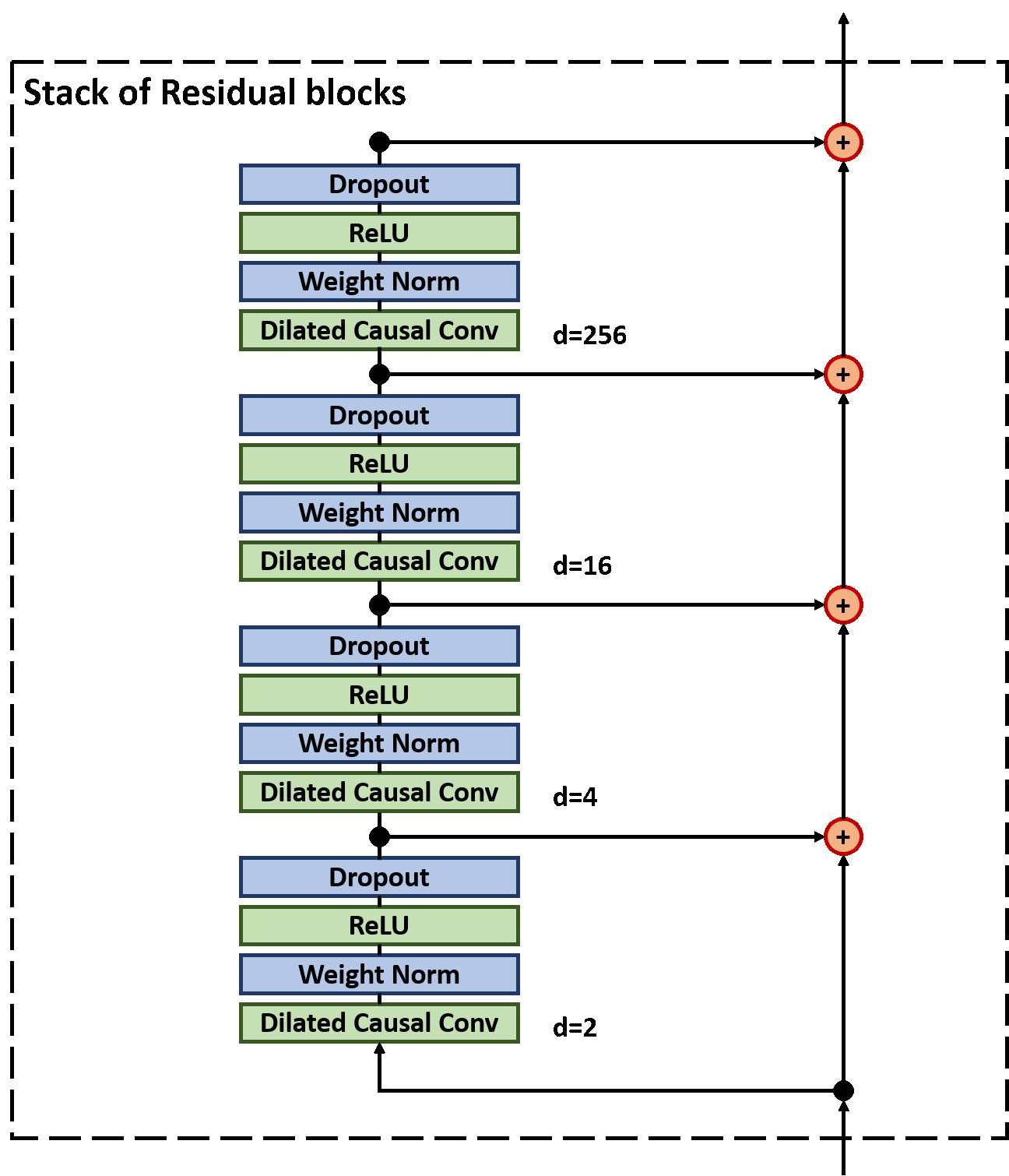}}
\caption[SeismicSimilarity's TCN Architecture]{A single stack of 4 dilated residual blocks commonly found in a Deep Temporal Convolutional Neural Network Architecture. In this case, the residual blocks have exponentially increasing dilation rates, increasing from 2 to 256 across the 4 blocks. This rapid dilation provides the network a wide receptive field which is critical for learning long-period features frequently found in time-series waveform data.}
\label{fig:TCN_stack_SeisSim}
\end{figure}

Using Eq. \eqref{eq:receptiveField_sim}, the TCN is designed to have an overall receptive field of 4171 samples (104 seconds), allowing it to learn long-period features down to 0.01 Hz, with just four dilated convolutional layers, as shown in Table \ref{tbl:TCN_layers_sim}. The TCN architecture consists of two residual stacks, shown in Fig.~\ref{fig:TCN_stack_SeisSim}, each with 50 filters and a kernel size (filter length) of 16 samples. Finally, the TCN output is encoded by a densely connected output layer with 32 nodes, and the final output vector is normalized to have unit length. This results in 553,835 trainable parameters, and a network which reduces the three-channel 21,600 dimensional input into just 32 dimensions, for a 99.9\% reduction in dimensionality.

\begin{table}[htbp]
\centering
\caption[SeismicSimilarity’s TCN Parameters]{TCN Layer Parameters}
\label{tbl:TCN_layers_sim}
\begin{tabular}{|c|c|c|c|}
\hline
\textbf{l} & \textbf{k} & \textbf{d} &\textbf{ Receptive Field} \\\hline
1        & 16          & 2        & 31              \\
2        & 16          & 4        & 91              \\
3        & 16          & 16       & 331             \\
4        & 16          & 256      & 4171       \\    
\hline
\end{tabular}
\end{table}

\subsection{Similarity Objective}

This embedding function is learned via a Triplet Network with batch-hard loss. Specifically, the batch size was set at 100, with $L$ (the number of distinct source events in a batch) and $K$ (the number of seismogram recordings for each event in a batch) both set equal to 10. In this way, each batch consists of 100 randomly selected seismograms, evenly represented across 10 different source events. These values were selected primarily on the basis of availability, since increasing $K$ beyond 10 would have limited the dataset (~95\% of the events in the USArray dataset were recorded by at least 10 stations), and increasing $L$ beyond 10 would require more memory than the 12Gb available in the Nvidia 1080Ti GPU used for training.  

Embedding space distances are computed using the $L_2$ Norm. Because the output of the embedding function is normalized, the embedding space vectors are all constrained to a hypersphere with radius = 1. This ensures a bounded distance between any two embeddings, as chord lengths are always bounded by [0,2] for any unit hypershpere. Because these pairwise distances are bounded, a fixed margin can be used throughout training~\citep{Sidiropoulos2014}.  In this work, the margin is fixed at $\alpha = 0.2$, which is common~\citep{Schroff2015}.

\subsection{Data Collection}

Learning a path-invariant measure for seismogram similarity requires a training dataset with many recordings of a single seismic event across many disparate paths. This is best accomplished by a dense network of seismometers across a wide region. EarthScope's USArray dataset is ideally suited for this endeavor. In particular, we utilize two EarthScope observatories, the Transportable Array and the Reference Array, as the basis for our Training and Test Sets, respectively.

The USArray Transportable Array (TA) consists of 400 temporary seismic instruments that were deployed at more than 2,000 temporary station locations across the Continental US between 2007 and 2015 ~\citep{Busby2018}. Each station utilized a broadband 3-channel (North-South, East-West and Vertical) instrument, installed in a post-hole configuration and digitized at 40 Hz. The instruments were generally one of three types, Guralp CMG3T, Quantera STS, or Nanometrics Trillium; the digitizers were primarily Kinemetrics Q330, Q680 or RefTek. In this work, our training and validation datasets are taken from the full array of TA seismograms, minus a random subset of 51 stations and a region of events located near the Rosebud mine in Montana, which were held out for testing. The training and validation sets were distinct in time, covering the periods from 2007-2013 and 2014, respectively. Associated arrival times were obtained by querying the ISC reviewed catalogs for any Continental US (CONUS) events over this period, resulting in 149,036 seismogram recordings of 4,825 distinct seismic events for the training set, and 22,561 seismogram recordings of 1,175 distinct seismic events for the validation set. A map detailing the layout of the training stations is shown in the left plot of Fig.~\ref{fig:SourceMap}. 

The USArray Reference Array (REF) consists of 120 permanent seismic instruments deployed across the Continental US, utilizing similar equipment as the Transportable Array. In this work, our test set is taken from the full array of TA and REF stations available from 2015 and 2016. Associated arrival times were obtained by querying the ISC reviewed catalog for CONUS events, resulting in a test set with 35,694 seismogram recordings of 2,452 distinct seismic events. All of the events in the test set are mutually exclusive with the training and validation data. Additionally, because of the stations and locations held out during testing, 6,934 of these seismograms were recorded by the 51 novel stations, and 87 seismograms represent events from the novel location near the Rosebud mine in Montana. Performance is evaluated explicitly on these novel data to explore the power and generalization of the technique.

\begin{figure}[htbp]
  \centering
  \setlength\fboxrule{0pt}
  \fbox{\includegraphics[width=3in]{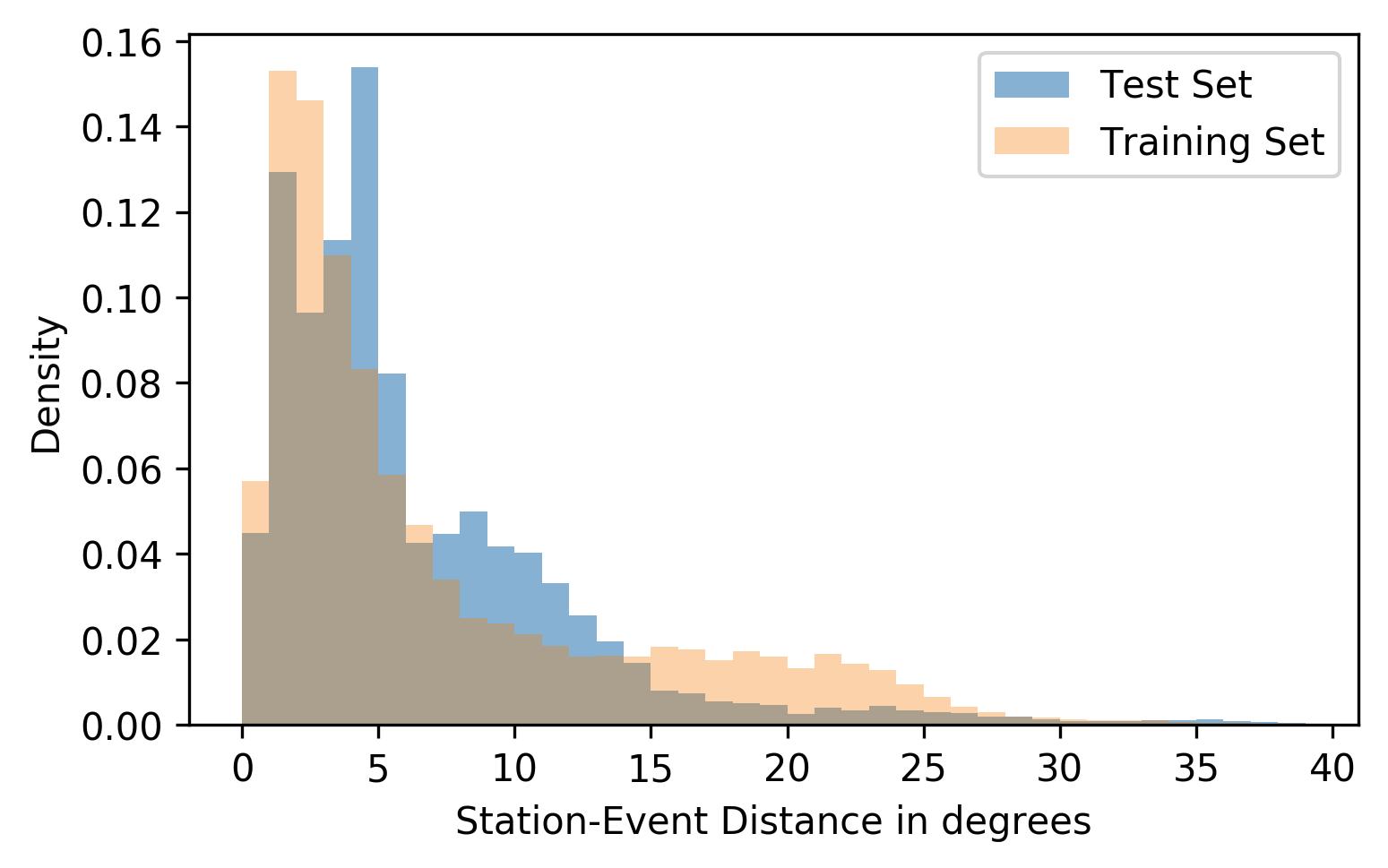}}
\caption[Distance Histograms]{Histogram showing the distributions of station-event distances for all seismograms in the test and training sets. The distributions show that the test and training sets are similar, and that the majority of the seismograms in the combined dataset were recorded within 15 degrees of the epicenter.}
\label{fig:dist_association}
\end{figure}

\begin{figure}[htbp]
  \centering
  \setlength\fboxrule{0pt}
  \fbox{\includegraphics[width=3in]{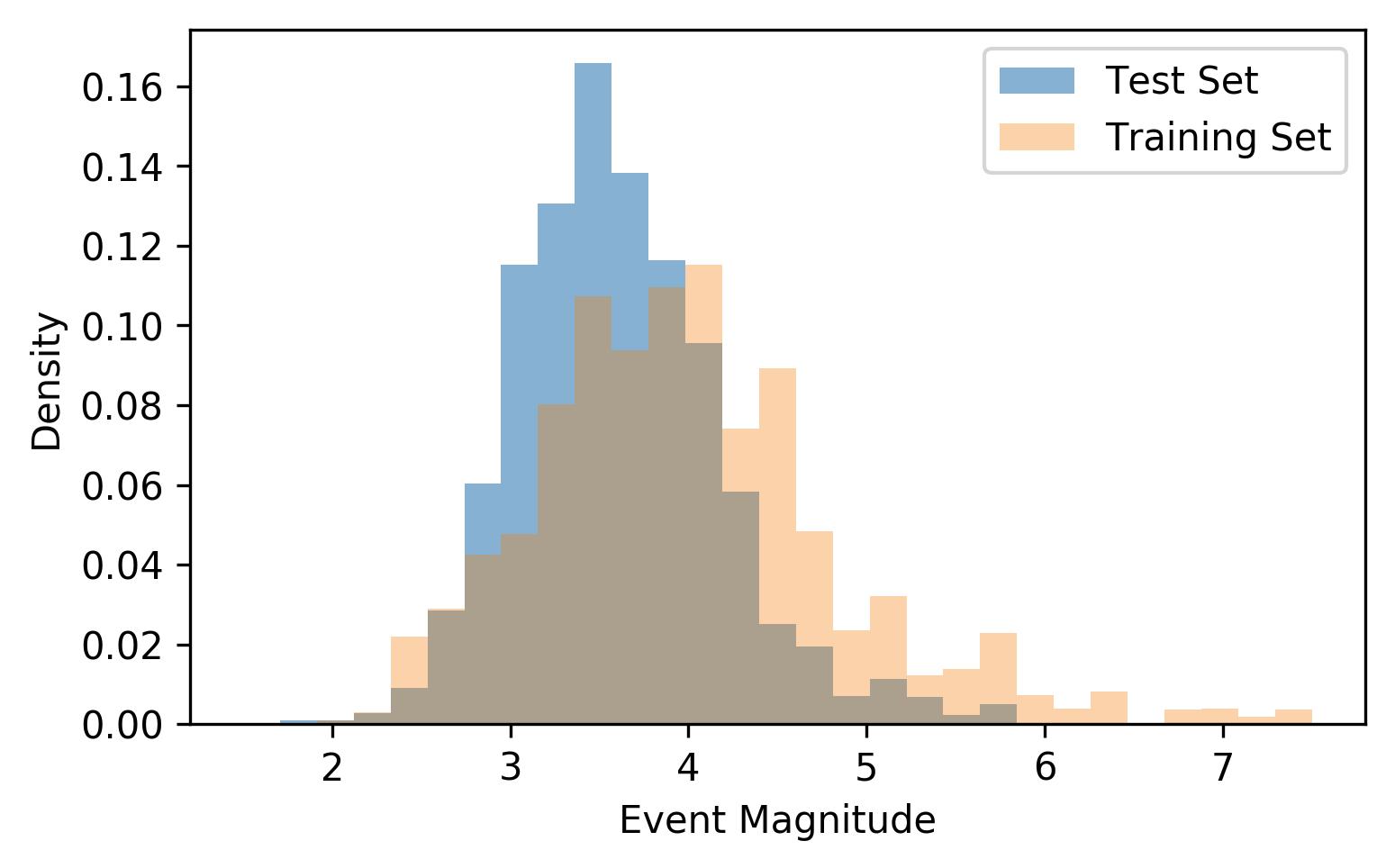}}
\caption[Magnitude Histograms]{Histogram showing the distributions of event magnitudes for all seismograms in the test and training sets. The distributions show that the test and training sets are similar, and that the majority of the events in the combined dataset have a magnitude between 2 and 5 Mb.}
\label{fig:mag_association}
\end{figure}

All three datasets, training, validation and test, were limited to events with near CONUS epicenters, as defined by the following limits on latitude and longitude: 25 $<$ LAT $<$ 50, and -125 $<$ LON $<$ -75. This accomplishes two purposes. First, this produces a catalog with more balanced samples of explosions and earthquakes, 207,291 and 26,568 respectively. Second, this restricts the study to regional signals. Regional signals are preferred due to the more manageable window length requirements vs teleseismic signals, as well as because the regional association task is much more interesting than the teleseismic association task; teleseismic signals recorded by such a dense regional network look much more similar even using traditional seismic similarity. We leave the exploration of this technique against teleseismic signals to future work. For completeness, we also have included histograms of seismogram station-to-event distances as well as event magnitudes for both the test and training sets, shown in Figs.~\ref{fig:dist_association} and~\ref{fig:mag_association}, respectively.

For each of the 207,291 seismograms in the combined datasets, a 180-second window is selected which includes the 30 seconds prior to the cataloged arrival time and the 150 seconds subsequent to the arrival. The only pre-processing applied to the raw data was a normalization and de-trending. This window size was chosen so as to ensure the presence of both P and S waves within the window. While, this long window does present the opportunity for multiple arrivals within a single window, investigation shows that this occurs in only 0.15\% of the seismograms in the dataset, and its effects are negligible on our results.

To create the training and validation triples, a generator function randomly selects an anchor, as well as positive (same event, different station) and negative (different) events. Due to multiple site recordings of many of the individual events (on average, each event was recorded by 30 different stations), there are upwards of 300 million possible triples, which makes this a robust training set for learning seismogram similarity.

\subsection{Evaluation Criteria}

To demonstrate the performance of the similarity measure, it is applied to two tasks: pairwise event association and source discrimination. Evaluation criteria for each of these tasks is shown below.

Event association is the process of correctly associating the arriving seismic phases of a single event across a network, and is a critical step in seismic analysis. The traditional algorithms used for this task have always been based on travel times and earth velocity models, however our method is similarity-based: we associate the seismograms entirely based on their pairwise similarity in the embedding space, with no external information about arrival times or recording locations. This is a binary classification task: given a pair of seismograms, $X_A$ and $X_B$, the algorithm must classify the pair as matched or unmatched, where a matched pair is defined as two seismogram recordings of the same event. Classification is accomplished by comparing the similarity-based test statistic, $S$, against a user defined threshold, $\tau$, as seen in Eq. \eqref{eq:association_task}.

\begin{equation}
\begin{aligned}
H_0& \text{: UNMATCHED ($X_A$ and $X_B$ depict distinct events)}\\
H_A& \text{: MATCHED ($X_A$ and $X_B$ depict a common event)}\\
& S = \frac{1}{\Big \langle f(X_A) , f(X_B) \Big \rangle} \\
& \text{reject $H_0$ if } S \geq \tau
 \label{eq:association_task}
\end{aligned}
\end{equation}

To report performance, a receiver operating characteristic (ROC) curve is built by varying $\tau$ across the full range of $S$, and plotting the rate of false positives against the rate of false negatives for each $\tau$. Additionally, for the threshold $\tau$ which maximizes accuracy, area under the ROC curve (AUC), binary classification accuracy, precision and recall are shown. The evaluation is performed across 50,000 random pairs of seismograms drawn from the test set, and compared directly against the results found in~\citep{McBrearty2019}. The results are also explored with respect to a subset of novel stations and events that were withheld during training, in order to better understand the abilities and limitations of the technique.

The source discrimination task is also formulated as binary classification, where unlabeled seismograms $X$ are classified as either explosion or earthquake, based on their embedding space similarities to both the centroid of a set of explosion templates, $X_{EXP}$ and the centroid of a set of earthquake templates, $X_{EQK}$. This is shown in Eq. \eqref{eq:discrimination_task}, where $\epsilon$ is machine precision. 

\begin{equation}
\begin{aligned}
H_0& \text{: EARTHQUAKE ($X$ depicts an earthquake)}\\
H_A& \text{: EXPLOSION ($X$ depicts an explosion)}\\
& S = \frac{\Big \langle f(X) , f(X_{EQK}) \Big \rangle}{\Big \langle f(X) , f(X_{EXP}) \Big \rangle + \epsilon} \\
& \text{reject $H_0$ if } S \geq \tau
 \label{eq:discrimination_task}
\end{aligned}
\end{equation}

The source discrimination test is performed against the full 35,694 seismograms in the test set.  The ROC curve, AUC, accuracy, precision, and recall are presented. 

Additionally, the performance of our similarity-based discriminator is directly compared to that of two state-of-the-art methods: the SVM-based discriminator proposed in~\citep{Kortstrom2016} and the SRSpec-CNN discriminator adapted from the work of~\citep{Nakano2019}. In particular, our SVM and CNN implementations both utilize the full 149,036 training waveforms from the training set. The SVM uses 36 features, composed of nine frequency bins ([1-3 Hz], [2-5 Hz], [4-7 Hz], [6-9 Hz], [8-11 Hz], [10-13 Hz], [12-15 Hz], [14-17 Hz], [16-19 Hz]) and four time divisions (P, P coda, S and S coda), with the S-P time differences based on the iasp91 velocity model. SRSpec-CNN uses 64x64 spectrogram images extracted from 180s normalized seismogram windows, with frequency bins between 2-10 Hz. 


\section{RESULTS}

\subsection{Pairwise Event Association}

\begin{figure}[htbp]
  \centering
  \setlength\fboxrule{0pt}
  \fbox{\includegraphics[width=3in]{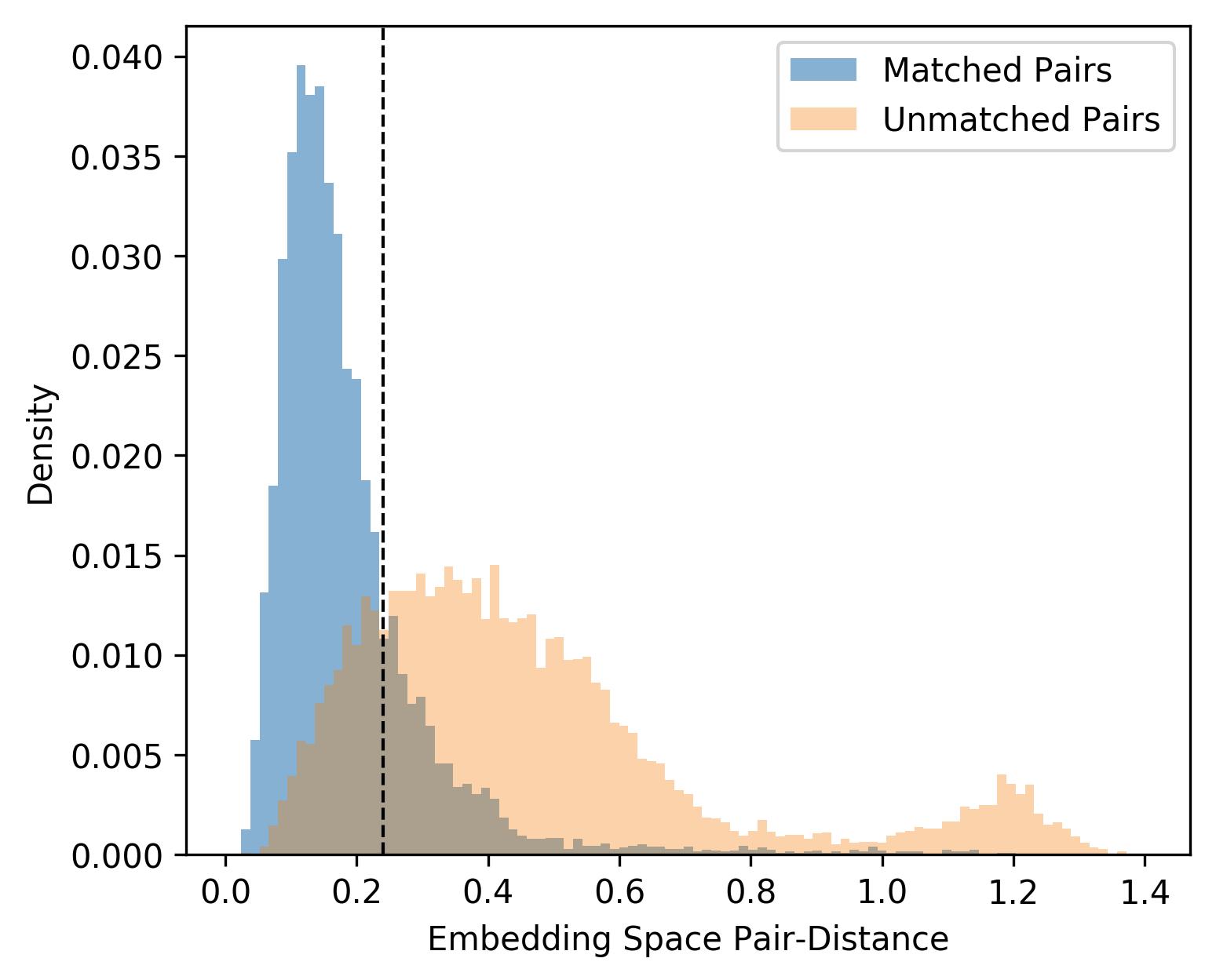}}
\caption[Event Association Histograms]{Histograms of matched and unmatched pair distances for the test set. The matched-pair distribution includes embedding space distances for 25,000 pairs of seismograms, where the two embeddings come from the same event. The unmatched-pair distribution includes embedding space distances for 25,000 pairs of seismograms, where the two embeddings come from different events. A cutoff threshold of 0.24 was used to obtain maximum classification accuracy, and is annotated by the dashed line. For this threshold, the area of overlap between the two density plots represents the total classification error, which is  $\sim$20\%.}
\label{fig:hist_association}
\end{figure}

To demonstrate that event association using this technique is possible, a special test set is created by sampling 50,000 pairs of seismograms from the test set, including 25,000 pairs of seismograms that originate from common events, and 25,000 pairs of seismograms that originate from different events. Plotting histograms of the embedding space distances for each pair, as shown in Fig.~\ref{fig:hist_association}, demonstrates that the distribution for matched-pair distances are considerably lower than the unmatched-pair distances.

\begin{figure}[htbp]
  \centering
  \setlength\fboxrule{0pt}
  \fbox{\includegraphics[width=3in]{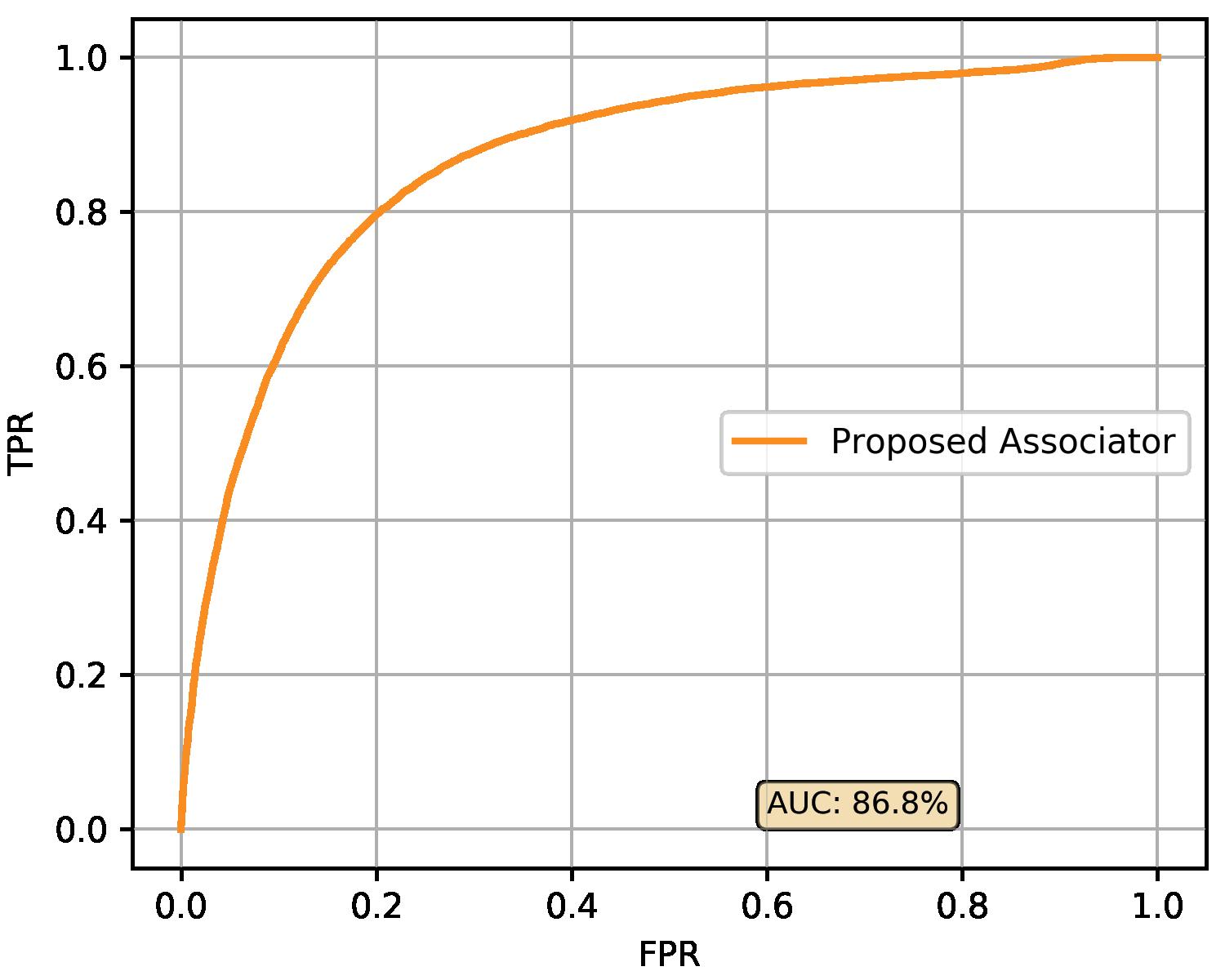}}
\caption[Event Association ROC Curve]{Receiver Operating Characteristic Curve for the Event Association task. The overall area under the curve is 86.8\%.}
\label{fig:ROC_association}
\end{figure}

We then apply the similarity-based association classifier defined in Eq. \eqref{eq:association_task}. The ROC curve for the task has an AUC of 86.8\% as shown in Fig.~\ref{fig:ROC_association}. The overall accuracy is 80.0\% with a precision and recall of 80.2\% and 79.6\%, respectively, and our results are nearly identical to the 80\% accuracy reported in~\citep{McBrearty2019}, extended across a much larger network of stations. Performance is also investigated with respect to the distance between recording stations. As noted previously, correlation-based seismogram similarity is known to decay exponentially with an increase in the distance between recording stations~\citep{Israelsson1990}. Our path-invariant measure is also negatively affected by increasing this distance, but the decay is only linear. This is clearly demonstrated in Table \ref{tbl:AssocDist}.

\begin{table}[htbp]
\centering
\caption[Waveform Association Performance vs Inter-station Distance]{Waveform Association Performance vs Inter-station Distance}
\label{tbl:AssocDist}
\begin{tabular}{|c|c|c|c|}
\hline
\textbf{Distance (km)} & \textbf{Precision} & \textbf{Recall} &\textbf{Accuracy} \\\hline
0000-0250 km	&   0.864	&   0.783	&   0.830 \\
0250-0500 km	&   0.852	&   0.791	&   0.827 \\
0500-0750 km	&   0.802	&   0.766	&   0.789 \\
0750-1000 km	&   0.805	&   0.789	&   0.799 \\
1000-1250 km	&   0.778	&   0.840	&   0.800 \\
1250-1500 km	&   0.785	&   0.811	&   0.794 \\
1500-1750 km	&   0.744	&   0.866	&   0.784 \\
1750-2000 km	&   0.731	&   0.863	&   0.773 \\
2000-2250 km	&   0.732	&   0.794	&   0.751 \\
2250-2500 km	&   0.741	&   0.826	&   0.769 \\
\hline
\end{tabular}
\end{table}

\begin{figure}[htbp]
  \centering
  \setlength\fboxrule{0pt}
  \fbox{\includegraphics[width=3in]{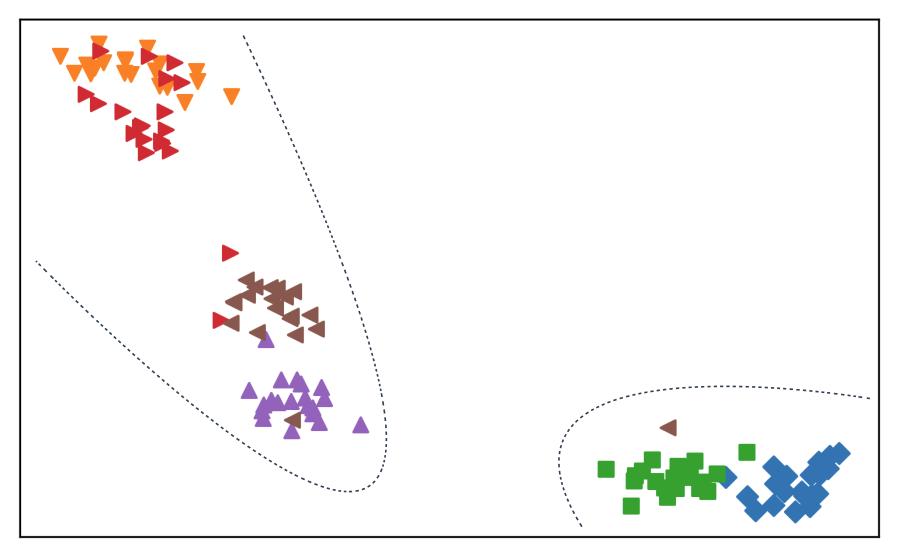}}
\caption[Event Association Embedding Space]{t-SNE Embeddings for Waveform Association. Six unique seismic events were randomly selected from the dataset, along with 20 seismograms for each event, recorded at various stations. These 120 seismograms were then mapped to the 32-dimensional embedding space via the trained neural network. Finally, the 32-dimensional embedding space was visualized here in two dimensions using t-SNE, with each unique event assigned a unique marker. The clustering of same-event embeddings is the result of shared feature commonalities between seismograms of that event. It is interesting to note that there appears to be some aggregate clustering as well, indicated by the dashed lines. This aggregate clustering is the result of feature commonalities shared across seismograms of multiple events. These inter-event commonalities are explored further, in our results for the source discrimination task.}
\label{fig:tSNE_association}
\end{figure}

To further investigate the ability of the embedding space to facilitate event association, Fig.~\ref{fig:tSNE_association}, displays 120 seismogram embeddings in 2-dimensions using t-Distributed Stochastic Neighbor Embedding (t-SNE)~\citep{Maaten2008}. The figure clearly demonstrates a clustering of embeddings of common events. However, there are obviously other clusters present as well, shown by the dashed lines in the plot. As it turns out, these other clusters can be quite useful, and are explored further in the discussion of the source discrimination task.

The ability of the embedding space to associate regional events across hundreds of stations with 80\% accuracy based entirely on waveform similarity is surprising, and begs the question: is the neural network really extracting generalized path-invariant features, or is it merely `memorizing' all the training paths exactly, in a way
that appears to support conclusions that are unwarranted. To answer this question, we investigate the ability of the embedding space to associate waveforms from novel stations and locations as detailed in Tables \ref{tbl:AssocNovelStation}, \ref{tbl:AssocNovelEvent} and \ref{tbl:AssocNovelAll}. Here, we find that although the performance does drop for such events, the drop is relatively minor. For instance, accuracy only drops from 80\% to 79\% when considering novel stations, which demonstrates that the neural network has indeed learned to extract features that are invariant to recording location, even novel ones. The accuracy drop is slightly more significant when considering novel event locations, decreasing from 80\% to 76\% for pairs where at least one event originated near the held-out Rosebud mine. This is understandable, as withholding training events from a certain source location obviously impairs the ability of the neural network to extract features unique to such events at test time.

\begin{table}[htbp]
\centering
\caption[SeismicSimilarity’s Association Performance for Novel Stations]{Association Performance for Novel Stations}
\label{tbl:AssocNovelStation}
\begin{tabular}{|c|c|c|c|}
\hline
\textbf{Novel STA?} & \textbf{COUNT} & \textbf{ERROR} &\textbf{ACCURACY} \\\hline
No 	&   32221 	&   6358 	&   0.80 \\
Yes	&   17779 	&   3712 	&   0.79 \\
\hline
\end{tabular}
\end{table}

\begin{table}[htbp]
\centering
\caption[SeismicSimilarity’s Association Performance for Novel Source Location]{Association Performance for Novel Source Location}
\label{tbl:AssocNovelEvent}
\begin{tabular}{|c|c|c|c|}
\hline
\textbf{Novel LOC?} & \textbf{COUNT} & \textbf{ERROR} &\textbf{ACCURACY} \\\hline
No 	&   49792 	&   10020 	&   0.80 \\
Yes	&   208 	&    50 	&   0.76 \\
\hline
\end{tabular}
\end{table}

\begin{table}[htbp]
\centering
\caption[SeismicSimilarity’s Association Performance for Novel Stations and Source Locations]{Association Performance for Novel Station and Location}
\label{tbl:AssocNovelAll}
\begin{tabular}{|c|c|c|c|}
\hline
\textbf{Novel STA\&LOC?} & \textbf{COUNT} & \textbf{ERROR} &\textbf{ACCURACY} \\\hline
No 	&   49908 	&   10038 	&   0.80 \\
Yes	&   92   	&   32  	&   0.65 \\
\hline
\end{tabular}
\end{table}

\subsection{Source Discrimination}

\begin{figure}[htbp]
  \centering
  \setlength\fboxrule{0pt}
  \fbox{\includegraphics[width=3in]{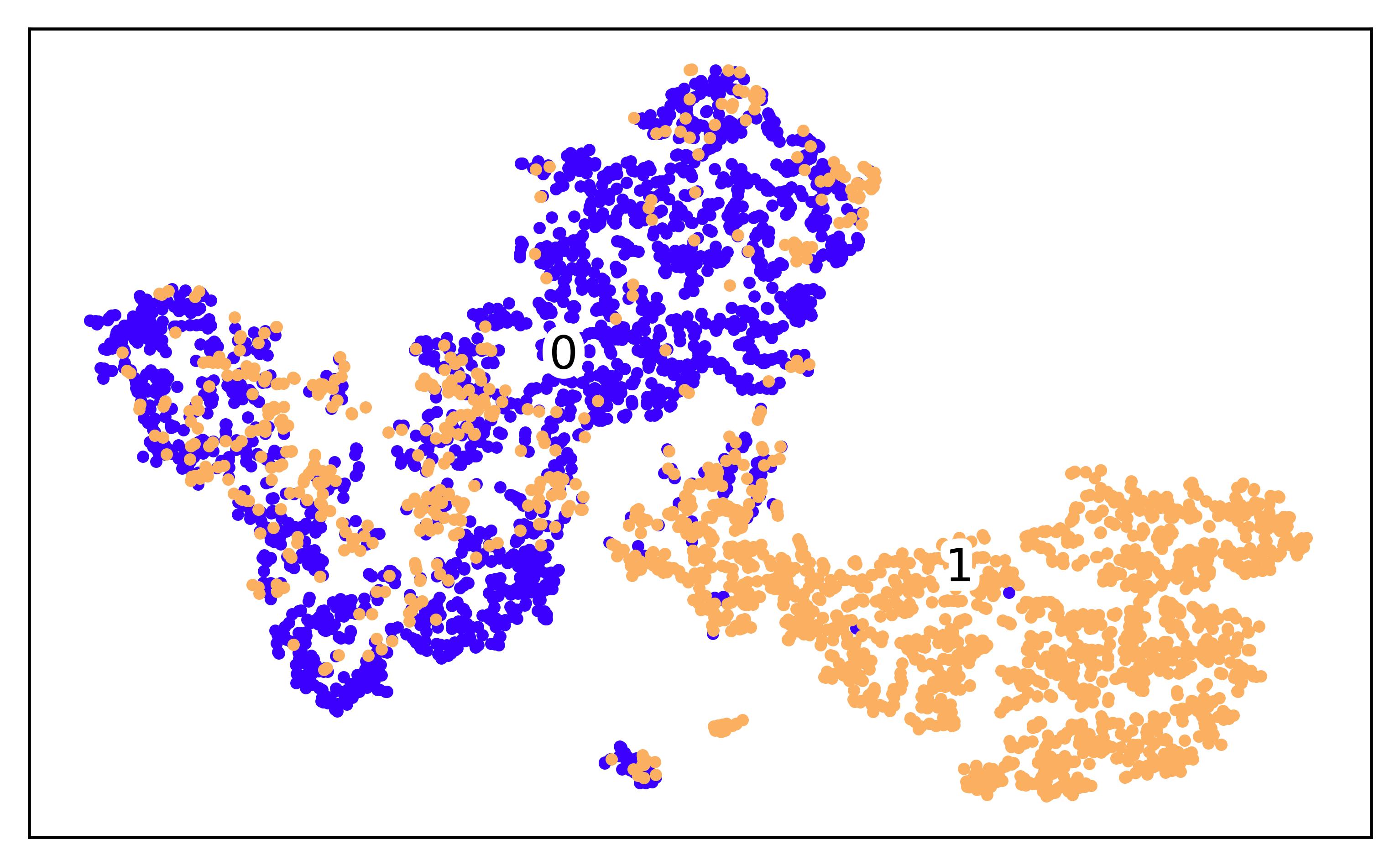}}
\caption[Source Discrimination Embedding Space]{Two-hundred embeddings are shown, visualized in 2D using t-SNE, and labeled according to source function. The light-colored dots represent explosions and the darker dots represent earthquakes; the cluster centroids are annotated by 1 and 0, respectively. The 2D clustering of embeddings demonstrates the inherent association between embeddings with a common source function.}
\label{fig:tSNE_discrimination}
\end{figure}

To further demonstrate the power of our embedding space, we consider its utility to facilitate template-based source discrimination. The results here are particularly interesting, as the neural network was not explicitly trained in this task: Although the neural network was exposed to many examples of earthquakes and explosions during training (207,291 and 26,568 respectively), the network had no access to these source labels. However, the network did have access to event labels, and was thus trained to extract features with source-specificity and path-invariance. Unsurprisingly, these source-specific features are well-suited for source discrimination. In Fig.~\ref{fig:tSNE_discrimination}, the embedding space is visualized using t-SNE, and labeled by source type, demonstrating a significant separation between the two source classes in the embedding space, with no pre-processing or training.

\begin{figure*}[htbp]
  \centering
  \setlength\fboxrule{0pt}
  \fbox{\includegraphics[width=6in]{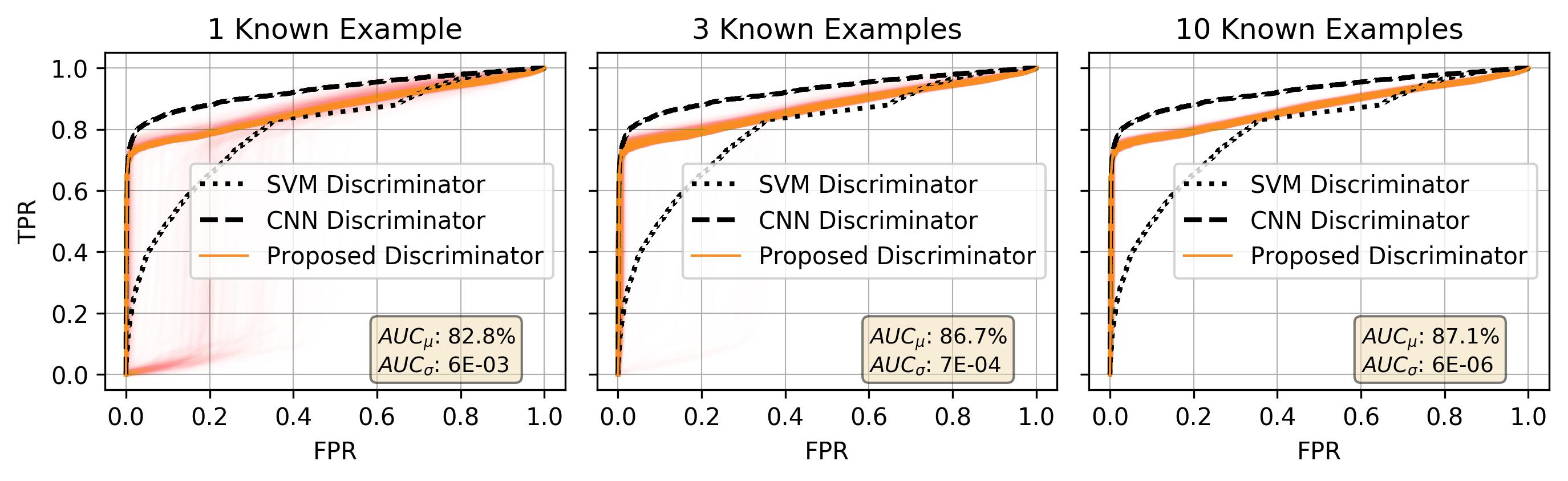}}
\caption[Source Discrimination ROC Curve]{Receiver Operating Characteristic Curves for the Source Discrimination task identifying all explosions. Three plots are shown, demonstrating performance across various numbers of templates (1, 3 and 10). Because the template are chosen randomly, we have performed 1,000 trials for each plot, with the results of each trial plotted as a separate curve. Performance converges nicely for only 3 templates. The dashed and dotted black lines show the performance of two alternative discriminators applied to the same dataset.}
\label{fig:ROC_discrimination}
\end{figure*}

\begin{figure*}[htbp]
  \centering
  \setlength\fboxrule{0pt}
  \fbox{\includegraphics[width=6in]{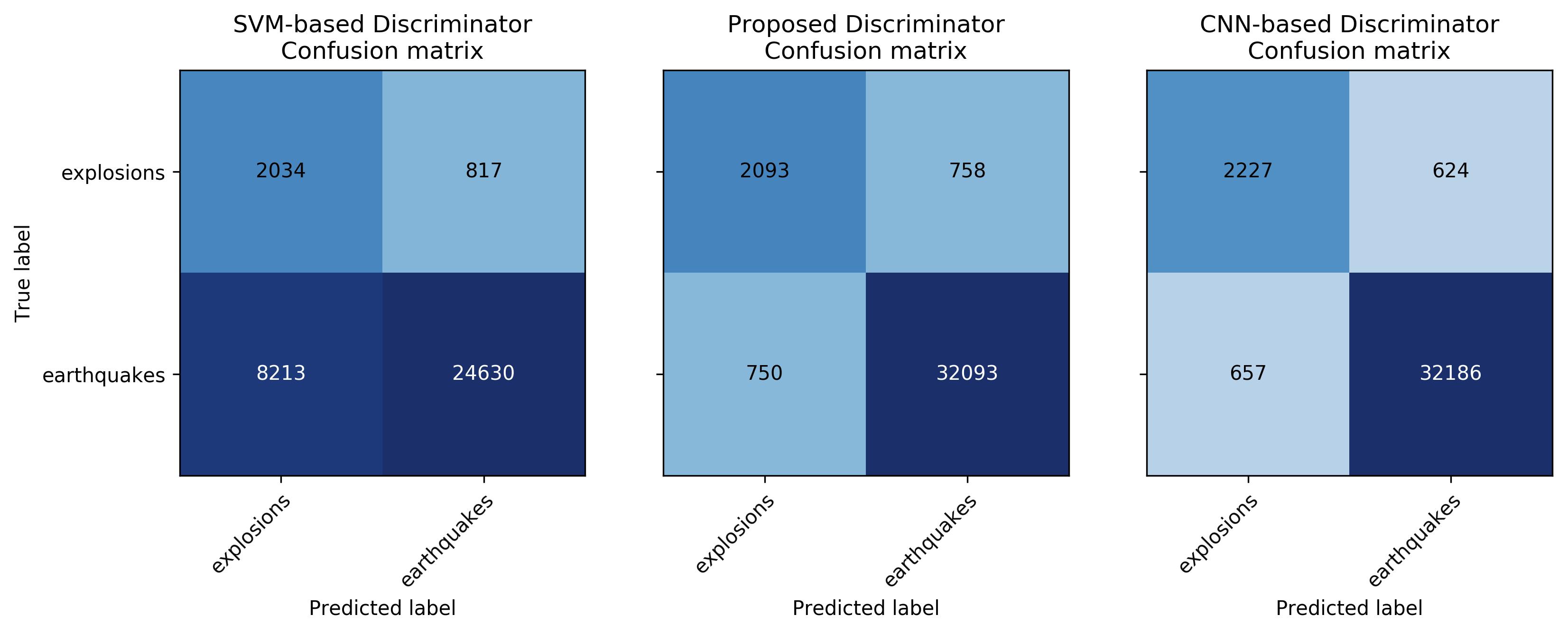}}
\caption[Source Discrimination Confusion Matrix]{Source Discrimination Confusion Matrix. Three matrices are shown, demonstrating performance of three source discrimination techniques against the test set. Our proposed Similarity-based discriminator utilizes a signal explosion template, whereas the SVM and CNN-based discriminators utilize a large training set with 10,000 labeled earthquakes and 10,000 labeled explosions.}
\label{fig:CM_discrimination}
\end{figure*}

Template-based discrimination performance is demonstrated with three different quantities of randomly-selected exemplar templates: 1, 3 and 10, as shown in Fig.~\ref{fig:ROC_discrimination}. The discriminator achieves a mean AUC of 82.8\% for just a single template. This is known as one-shot learning, and enables the creation of a viable classification algorithm with only a single training example. The variance on this AUC is a bit high; however with three templates, this method achieves an AUC of 86.7\% with low variance. Choosing the threshold so as to maximize accuracy, the algorithm is then evaluated for accuracy, precision and recall, which are recorded at 95.8\%, 73.4\% and 73.6\% respectively, which exceeds the performance of the SVM discriminator, but falls just short of the 96.4\%, 78.1\% and 77.2\% performance achieved by the SRSpec-CNN classifier applied to the same dataset, as detailed in Fig.~\ref{fig:CM_discrimination}. This minimal performance gap between SRSpec-CNN and our template-based discriminator is surprising, given that SRSpec-CNN is a state-of-the-art fully-supervised method with well-engineered features while our template-based discriminator utilizes semi-supervised learning, with access to just a single template.

\subsection{Computation Time}

Optimizing the Neural Network during training requires considerable computation time: approximately 30 hours on the aforementioned Nvidia 1080Ti. However the model only needs to be trained once; after training, deployment is quite fast at runtime, requiring only 1.8 milliseconds to transform a single 180 second window of 3-channel waveform data onto the embedding space. This represents a four-fold improvement over the 7.6 milliseconds required to take the same waveform and extract the spectrogram features used in traditional source discrimination. Runtimes for the Validation and Test sets are shown in Table \ref{tbl:RunTimes}.

\begin{table}[htbp]
\centering
\caption[SeismicSimilarity’s Computation Time]{Comparison of Runtimes against the Validation and Test Sets}
\label{tbl:RunTimes}
\begin{tabular}{|c|c|c|c|}
\hline
 & \textbf{Val Set} & \textbf{Test Set} &\textbf{Runtime} \\
 & \textbf{(22,561 samps)} & \textbf{(35,694 samps)} &\textbf{per samp} \\
\hline
NN Embeddings 	        &   41s 	&   66s 	&   1.8ms \\
Spectral Features    &   171s   	&   271s  	&   7.6ms \\
\hline
\end{tabular}
\end{table}

\section{CONCLUSION}

To date, almost all seismogram similarity measures have been based on the cross-correlation function, constraining them to relatively path-dominant similarity, and limiting their use to repetitive and geographically localized signals. In this work, we have presented a path-invariant measure for seismogram similarity, based on a deep triplet network architecture. We have demonstrated the effectiveness of this measure for both pairwise event association and template-based source discrimination.

For the pairwise association task, our similarity measure is able to achieve an accuracy of 80\%, without any knowledge of recording time or phase type, across a large and diverse regional network. This is a significant advancement on the work done by McBrearty~\citep{McBrearty2019}, both in terms of providing increased generalization and extended path distances. And while pairwise-similarity is certainly a weaker evidence for association than a standard moveout curve, it does present a viable complimentary validation tool, which could be used to augment existing methods of automatic association. For instance, given an event list from an automatic associator, each event can be scored based on its embedding-space distance from the cluster centroid, and dissimilar events can simply be rejected or flagged for further analyst review based on the desired type-I error rate. Our future work involves constructing a more robust framework for this task, using additional layers of machine learning.

The results for template-based source discrimination are also quite promising. The 95.8\% classification accuracy achieved for explosion discrimination is impressive in its own right. However it is astounding considering that the discrimination is based on a single template waveform. This result is not only useful for identifying explosions, but also holds considerable promise for other discrimination tasks. In fact, as with most semi-supervised techniques, the real potential of our similarity-based classifier lies in its application to less well-studied and less robustly labeled classes. For instance, while the USGS CONUS catalog used in this work includes painstakingly labeled explosions, such labels are simply not available for many other regions. Similarly, there are numerous other source types of interest (volcanoes, ice quakes, rock bursts, tremors, ripple-fire blasts, etc) for which labels may be scarce or unavailable. As such, our method holds considerable potential for training future discriminators on less well-studied source functions, especially when training examples are limited and fully-supervised methods are unavailable. 

In conclusion, we believe that the findings in this work represent an important step forward in the field of seismogram similarity, demonstrating that such similarity measures do not need to be constrained to the path-dominant correlation-based detectors traditionally implemented. However, there is still much work to be done, especially in the application of this method across more diverse datasets, including global networks and teleseismic signals.

\section{Data and Resources}

The raw seismograms used in this study were collected as part of Earth Scope's USArray experiment~\citep{Busby2018}, and can be accessed via the Incorporated Research Institutions for Seismology (IRIS) Database using ObsPy~\citep{Beyreuther2010}.

Arrival-time catalogs for each station were downloaded through a web query of the International Seismological Centre (ISC)  Bulletin  for  seismic  arrivals:

\mbox{\url{http://www.isc.ac.uk/iscbulletin/search/arrivals/}} (last accessed February 2019).

The Neural Network Architecture was implemented in Keras~\citep{chollet2015}, using the keras-tcn python package written by Philippe R\'emy:

\mbox{\url{https://github.com/philipperemy/keras-tcn}} (last accessed February 2019).

The batch-hard algorithm was implemented in Tensorflow~\citep{tensorflow2015-whitepaper}, and adapted from the work of Olivier Moindrot, which can be found at:

\mbox{\url{https://omoindrot.github.io/triplet-loss}} (last accessed February 2019).

A repository containing the code and trained models described in this manuscript has been made available on github, and can be found at: 

\mbox{\url{https://github.com/joshuadickey/seis-sim}} (last accessed September 2019).

\section{Acknowledgment}
The results presented in this paper are solely the opinion of the authors; they do not represent the official position or policy of the United~States~Government.


\begin{thebibliography}{}

\bibitem[Abadi et~al., 2015]{tensorflow2015-whitepaper}
Abadi, M., Agarwal, A., Barham, P., Brevdo, E., Chen, Z., Citro, C., Corrado,
  G.~S., Davis, A., Dean, J., Devin, M., Ghemawat, S., Goodfellow, I., Harp,
  A., Irving, G., Isard, M., Jia, Y., Jozefowicz, R., Kaiser, L., Kudlur, M.,
  Levenberg, J., Man{\'{e}}, D., Monga, R., Moore, S., Murray, D., Olah, C.,
  Schuster, M., Shlens, J., Steiner, B., Sutskever, I., Talwar, K., Tucker, P.,
  Vanhoucke, V., Vasudevan, V., Vi{\'{e}}gas, F., Vinyals, O., Warden, P.,
  Wattenberg, M., Wicke, M., Yu, Y., and Zheng, X. (2015).
\newblock {TensorFlow: Large-Scale Machine Learning on Heterogeneous
  Systems}.

\bibitem[Bai et~al., 2018]{Bai2018}
Bai, S., Kolter, J.~Z., and Koltun, V. (2018).
\newblock {An Empirical Evaluation of Generic Convolutional and Recurrent
  Networks for Sequence Modeling}.
\newblock {\em arXiv e-prints}, abs/1803.0.

\bibitem[Beauc{\'{e}} et~al., 2017]{Beauce2017}
Beauc{\'{e}}, E., Frank, W.~B., and Romanenko, A. (2017).
\newblock {Fast Matched Filter (FMF): An Efficient Seismic Matched‐Filter
  Search for Both CPU and GPU Architectures}.
\newblock {\em Seismological Research Letters}, 89(1):165--172.

\bibitem[Belkin and Niyogi, 2003]{Belkin2003}
Belkin, M. and Niyogi, P. (2003).
\newblock {Laplacian Eigenmaps for Dimensionality Reduction and Data
  Representation}.
\newblock {\em Neural Computation}, 15(6):1373--1396.

\bibitem[Bergen and Beroza, 2018a]{Bergen2018a}
Bergen, K.~J. and Beroza, G.~C. (2018a).
\newblock {Detecting earthquakes over a seismic network using single-station
  similarity measures}.
\newblock {\em Geophysical Journal International}, 213(3):1984--1998.

\bibitem[Bergen and Beroza, 2018b]{Bergen2018}
Bergen, K.~J. and Beroza, G.~C. (2018b).
\newblock {Earthquake Fingerprints: Extracting Waveform Features for
  Similarity-Based Earthquake Detection}.
\newblock {\em Pure and Applied Geophysics}.

\bibitem[Beyreuther et~al., 2010]{Beyreuther2010}
Beyreuther, M., Barsch, R., Krischer, L., Megies, T., Behr, Y., and Wassermann,
  J. (2010).
\newblock {ObsPy: A Python Toolbox for Seismology}.
\newblock {\em Seismological Research Letters}, 81(3):530--533.

\bibitem[Bormann and {IASPEI}, 2012]{Bormann2002a}
Bormann, P. and {IASPEI} (2012).
\newblock {\em {New Manual of Seismological Observatory Practice (NMSOP-2)}},
  volume 2 Volumes.
\newblock GFZ German Research Centre for Geosciences, Potsdam, DE, 2 edition.

\bibitem[Burges et~al., 2003]{Burges2003}
Burges, C. J.~C., Platt, J.~C., and Jana, S. (2003).
\newblock {Distortion discriminant analysis for audio fingerprinting}.
\newblock {\em IEEE Transactions on Speech and Audio Processing},
  11(3):165--174.

\bibitem[Busby et~al., 2018]{Busby2018}
Busby, R., Woodward, R., Hafner, K., Vernon, F., and Frassetto, A. (2018).
\newblock {The Design and Implementation of EarthScope’s USArray
  Transportable Array in the Conterminous United States and Southern Canada}.
\newblock Technical report, Earth Scope.

\bibitem[C.~Pechmann and Kanamori, 1982]{Pechmann1982}
C.~Pechmann, J. and Kanamori, H. (1982).
\newblock {Waveforms and spectra of preshocks and aftershocks of the 1979
  Imperial Valley, California, Earthquake: evidence for fault hetergeneity}.
\newblock {\em Journal of Geophysical Research}, 871:10579--10598.

\bibitem[Chen et~al., 2009]{Chen2009}
Chen, Y., Garcia, E.~K., Gupta, M.~R., Rahimi, A., and Cazzanti, L. (2009).
\newblock {Similarity-based Classification: Concepts and Algorithms}.
\newblock {\em J. Mach. Learn. Res.}, 10:747--776.

\bibitem[Chollet and {others}, 2015]{chollet2015}
Chollet, F. and {others} (2015).
\newblock {Keras}.
\newblock {\textbackslash}url{\{}https://keras.io{\}}.

\bibitem[Chopra et~al., 2005]{Chopra2005b}
Chopra, S., Hadsell, R., and LeCun, Y. (2005).
\newblock {Learning a Similarity Metric Discriminatively, with Application to
  Face Verification}.
\newblock {\em Proceedings of the IEEE Computer Society Conference on Computer
  Vision and Pattern Recognition}, 1:539--546.

\bibitem[Dickey et~al., 2019]{Dickey2019}
Dickey, J., Borghetti, B., and Junek, W. (2019).
\newblock {Improving Regional and Teleseismic Detection for Single-Trace
  Waveforms Using a Deep Temporal Convolutional Neural Network Trained with an
  Array-Beam Catalog}.
\newblock {\em Sensors}, 19(3).

\bibitem[Dodge and Walter, 2015]{Dodge2015}
Dodge, D.~A. and Walter, W.~R. (2015).
\newblock {Initial Global Seismic Cross‐Correlation Results: Implications for
  Empirical Signal Detectors}.
\newblock {\em Bulletin of the Seismological Society of America},
  105(1):240--256.

\bibitem[Dysart and Pulli, 1987]{Dysart1987}
Dysart, P.~S. and Pulli, J.~J. (1987).
\newblock {Spectral study of regional earthquakes and chemical explosions
  recorded at the NORESS array}.
\newblock Technical report, Center for Seismic Studies.

\bibitem[Frankel, 1982]{Frankel1982}
Frankel, A. (1982).
\newblock {Precursors to a magnitude 4.8 earthquake in the Virgin Islands:
  Spatial clustering of small earthquakes, anomalous focal mechanisms, and
  earthquake doublets}.
\newblock {\em Bulletin of the Seismological Society of America},
  72(4):1277--1294.

\bibitem[Giannakis and Tsatsanis, 1990]{Giannakis1990}
Giannakis, G.~B. and Tsatsanis, M.~K. (1990).
\newblock {Signal detection and classification using matched filtering and
  higher order statistics}.
\newblock {\em IEEE Transactions on Acoustics, Speech, and Signal Processing},
  38(7):1284--1296.

\bibitem[Gibbons and Ringdal, 2006]{Gibbons2006}
Gibbons, S.~J. and Ringdal, F. (2006).
\newblock {The detection of low magnitude seismic events using array-based
  waveform correlation}.
\newblock {\em Geophysical Journal International}, 165(1):149--166.

\bibitem[Hadsell et~al., 2006]{Hadsell2006b}
Hadsell, R., Chopra, S., and LeCun, Y. (2006).
\newblock {Dimensionality reduction by learning an invariant mapping}.
\newblock {\em Proceedings of the IEEE Computer Society Conference on Computer
  Vision and Pattern Recognition}, 2:1735--1742.

\bibitem[Harris, 1991]{Harris1991}
Harris, D.~B. (1991).
\newblock {A waveform correlation method for identifying quarry explosions}.
\newblock {\em Bulletin of the Seismological Society of America},
  81(6):2395--2418.

\bibitem[Harris, 2006]{Harris2006}
Harris, D.~B. (2006).
\newblock {Subspace Detectors: Theory}.
\newblock Technical report, Lawrence Livermore National Laboratory (LLNL),
  Livermore, CA.

\bibitem[Hermans et~al., 2017]{Hermans2017}
Hermans, A., Beyer, L., and Leibe, B. (2017).
\newblock {In Defense of the Triplet Loss for Person Re-Identification}.
\newblock {\em arXiv e-prints}, abs/1703.0.

\bibitem[Hoffer and Ailon, 2015]{Hoffer2015b}
Hoffer, E. and Ailon, N. (2015).
\newblock {Deep metric learning using triplet network}.
\newblock {\em Lecture Notes in Computer Science (including subseries Lecture
  Notes in Artificial Intelligence and Lecture Notes in Bioinformatics)},
  9370(1271):84--92.

\bibitem[Hutchings and Wu, 1990]{Hutchings1990}
Hutchings, L. and Wu, F. (1990).
\newblock {Empirical Green's Functions from small earthquakes: A waveform study
  of locally recorded aftershocks of the 1971 San Fernando Earthquake}.
\newblock {\em Journal of Geophysical Research}, 95:1187--1214.

\bibitem[Israelsson, 1990]{Israelsson1990}
Israelsson, H. (1990).
\newblock {Correlation of waveforms from closely spaced regional events}.
\newblock {\em Bulletin of the Seismological Society of America},
  80(6B):2177--2193.

\bibitem[Jain et~al., 2008]{Jain2008}
Jain, P., Kulis, B., Dhillon, I.~S., and Grauman, K. (2008).
\newblock {Online metric learning and fast similarity search}.
\newblock {\em Advances Neural Information Processing Systems}, pages 1--8.

\bibitem[Jain et~al., 2009]{Jain2009}
Jain, P., Kulis, B., V.~Davis, J., and S.~Dhillon, I. (2009).
\newblock {Metric and Kernel Learning Using a Linear Transformation}.
\newblock {\em Journal of Machine Learning Research}, 13.

\bibitem[Jang and Yoo, 2009]{Jang2009}
Jang, D. and Yoo, C.~D. (2009).
\newblock {Fingerprint matching based on distance metric learning}.
\newblock In {\em 2009 IEEE International Conference on Acoustics, Speech and
  Signal Processing}, pages 1529--1532.

\bibitem[Kanamori and Ishida, 1978]{Kanamori1978}
Kanamori, H. and Ishida, M. (1978).
\newblock {The foreshock activity of the 1971 San Fernando earthquake,
  California}.
\newblock {\em Bulletin of the Seismological Society of America},
  68(5):1265--1279.

\bibitem[Koch et~al., 2015]{Koch2015}
Koch, G., Zemel, R., and Salakhutdinov, R. (2015).
\newblock {Siamese neural networks for one-shot image recognition}.
\newblock In {\em ICML Deep Learning Workshop}, volume~2.

\bibitem[Kong et~al., 2018]{Kong2018}
Kong, Q., Trugman, D.~T., Ross, Z.~E., Bianco, M.~J., Gerstoft, P., and Meade,
  B.~J. (2018).
\newblock {Machine Learning in Seismology: Turning Data into Insights}.
\newblock {\em Seismological Research Letters}, 90(1):3--14.

\bibitem[Kortström et~al., 2016]{Kortstrom2016}
Kortström, J., Uski, M., Tiira, T. (2016).
\newblock {Automatic classification of seismic events within a regional seismograph network}.
\newblock {\em Computers and Geosciences}, 87(1):22-30.

\bibitem[Kumar et~al., 2016]{Kumar2016}
Kumar, V. B.~G., Carneiro, G., and Reid, I. (2016).
\newblock {Learning Local Image Descriptors with Deep Siamese and Triplet
  Convolutional Networks by Minimizing Global Loss Functions}.
\newblock In {\em IEEE Conference on Computer Vision and Pattern Recognition
  (CVPR)}, pages 5385--5394.

\bibitem[Leal-Taixe et~al., 2016]{Leal-Taixe2016b}
Leal-Taixe, L., Canton-Ferrer, C., and Schindler, K. (2016).
\newblock {Learning by Tracking: Siamese CNN for Robust Target Association}.
\newblock {\em IEEE Computer Society Conference on Computer Vision and Pattern
  Recognition Workshops}, pages 418--425.

\bibitem[LeCun et~al., 1989]{Lecun1989}
LeCun, Y., Boser, B., Denker, J.~S., Henderson, D., Howard, R.~E., Hubbard, W.,
  and Jackel, L.~D. (1989).
\newblock {Backpropagation Applied to Handwritten Zip Code Recognition}.
\newblock {\em Neural Computation}, 1(4):541--551.

\bibitem[Maaten and Hinton, 2008]{Maaten2008}
Maaten, L. v.~d. and Hinton, G. (2008).
\newblock {Visualizing Data using t-SNE}.
\newblock {\em Journal of Machine Learning Research}, 9(Nov):2579--2605.

\bibitem[McBrearty et~al., 2019]{McBrearty2019}
McBrearty, I.~W., Delorey, A.~A., and Johnson, P.~A. (2019).
\newblock {Pairwise Association of Seismic Arrivals with Convolutional Neural
  Networks}.
\newblock {\em Seismological Research Letters}, 90(2).

\bibitem[Motoya and Abe, 1985]{Motoya1985}
Motoya, Y. and Abe, K. (1985).
\newblock {Waveform Similarity among Foreshocks and Aftershocks of the October
  18, 1981, Eniwa, Hokkaido, Earthquake}.
\newblock In Kisslinger, C. and Rikitake, T., editors, {\em Practical
  Approaches to Earthquake Prediction and Warning}, pages 627--636. Springer
  Netherlands, Dordrecht.

\bibitem[Nakano et~al., 2019]{Nakano2019}
Nakano, M., Sugiyama, D., Hori, T., Kuwatani, T., and Tsuboi, S. (2019).
\newblock {Discrimination of Seismic Signals from Earthquakes and Tectonic Tremor by Applying a Convolutional Neural Network to Running Spectral Images}.
\newblock {\em Seismological Research Letters}, 90(2A):530--538.

\bibitem[Schroff et~al., 2015]{Schroff2015}
Schroff, F., Kalenichenko, D., and Philbin, J. (2015).
\newblock {FaceNet: A Unified Embedding for Face Recognition and Clustering}.
\newblock In {\em CVPR}.

\bibitem[Schulte-Theis and Joswig, 1993]{Schulte1993}
Schulte-Theis, H. and Joswig, M. (1993).
\newblock {Master-event correlations of weak local earthquakes by dynamic
  waveform matching}.
\newblock {\em Geophysical Journal International}, 113(3):562--574.

\bibitem[Sidiropoulos, 2014]{Sidiropoulos2014}
Sidiropoulos, P. (2014).
\newblock {N-sphere chord length distribution}.
\newblock {\em arXiv preprint}.

\bibitem[Stauder and Ryall, 1967]{Stauder1967}
Stauder, W. and Ryall, A. (1967).
\newblock {Spatial distribution and source mechanism of microearthquakes in
  Central Nevada}.
\newblock {\em Bulletin of the Seismological Society of America},
  57(6):1317--1345.
  
\bibitem[Tibi et~al., 2017]{Tibi2017}
Tibi, R., Young, C., Gonzales, A., Ballard, S., Encarnacao, A. (2017).
\newblock {Rapid and robust cross-correlation-based seismic signal identification using an approximate nearest neighbor method}.
\newblock {\em Bulletin of the Seismological Society of America},
  107(4):1954-1968.
  

\bibitem[Waldhauser and Schaff, 2008]{Waldhauser2008}
Waldhauser, F. and Schaff, D. (2008).
\newblock {Large-scale relocation of two decades of Northern California seismicity using cross-correlation and double-difference methods}.
\newblock {\em Journal of Geophysical Research: Solid Earth},
  113(B8):0148-0227.
  
\bibitem[Wang et~al., 2014]{Wang2014}
Wang, J., Song, Y., Leung, T., Rosenberg, C., Wang, J., Philbin, J., Chen, B.,
  and Wu, Y. (2014).
\newblock {Learning Fine-grained Image Similarity with Deep Ranking}.
\newblock {\em arXiv e-prints}, page arXiv:1404.4661.

\bibitem[Xing et~al., 2002]{Xing2002}
Xing, E.~P., Jordan, M.~I., Russell, S., Ng, A.~Y., Jordan, M.~I., and Russell,
  S. (2002).
\newblock {Distance metric learning with application to clustering with
  side-information}.
\newblock {\em Advances in neural information processing systems (NIPS)},
  15(2):505--512.

\bibitem[Yoon et~al., 2015]{Yoon2015}
Yoon, C.~E., O'Reilly, O., Bergen, K.~J., and Beroza, G.~C. (2015).
\newblock {Earthquake detection through computationally efficient similarity
  search}.
\newblock {\em Science advances}, 1(11):e1501057--e1501057.

\bibitem[Zhang and Wen, 2015]{Zhang2015}
Zhang, M. and Wen, L. (2015).
\newblock {An effective method for small event detection: match and locate
  (MnL)}.
\newblock {\em Geophysical Journal International}, 200(3):1523--1537.

\end{thebibliography}

\end{document}